
\magnification=1200
\baselineskip=12pt

\bigskip

\centerline{{\bf SPIN CHAINS WITH TWISTED MONODROMY}
\footnote*
{Work supported in part by the National Science Foundation under grants
DMS-01-04261 and DMS-98-00783, \ $^{\dagger}$ on leave Landau Institute for
Theoretical Physics and ITEP, Moscow, Russia}}

\bigskip

\centerline{I.M. Krichever$^{\dagger}$ and D.H. Phong$^{\ddagger}$}

\bigskip
\centerline{Department of Mathematics}
\centerline{Columbia University}
\centerline{New York, NY 10027}
\centerline{$^{\dagger}$email: krichev@math.columbia.edu}
\centerline{$^{\ddagger}$email: phong@math.columbia.edu}
\bigskip
\centerline{}
\bigskip
\centerline{}
\bigskip
\centerline{\bf Abstract}

The integrable model corresponding to the ${\cal N}=2$ supersymmetric SU(N)
gauge theory with matter in the symmetric representation is constructed.
It is a spin chain model, whose key feature is a new twisted monodromy
condition.

\vfill\break

\centerline{\bf I. INTRODUCTION}

\bigskip

The 1994 work of Seiberg and Witten [14,15] revealed the existence of
a deep correspondence between supersymmetric gauge theories and integrable models [5,7,12,13]. 
However, the specific
list of correspondences is still far from complete at the present
time (c.f. [1-4] and references therein). In particular, it is still
not known how to construct the integrable model
which corresponds to the ${\cal N}=2$ $SU(N)$ gauge theory with
matter in the symmetric representation, although the spectral curve
has been identified by Landsteiner and Lopez using M Theory [11,6].

\medskip

The purpose of this paper is solve this problem.
In [10], we had solved a similar, but simpler problem,
which is to construct the integrable model corresponding to the 
$SU(N)$ gauge theory with matter in the antisymmetric representation.
The main new difficulty is the asymmetry between
the orders of the zero and the pole of the eigenvalues of the
monodromy operator at the two compactification points of the
spectral curve. The desired integrable model turns out to be
still a spin chain $p_n,q_n$, but whose main
feature is a new periodicity condition linking $p_{n+N+2},q_{n+N+2}$ to $p_n,q_n$ 
through a twisted 
monodromy operator. Such periodicity conditions have not
appeared
before in the literature, and we take the opportunity to discuss them in some
detail. An earlier proposal of how they can be used to construct other
integrable models is in [8], but not pursued further there. 

\medskip
The mathematical problem can be formulated very simply. It is to find
an integrable Hamiltonian system with spectral parameter $x$, spectral
curve
$$
R(x,y)\equiv y^3+f_N(x)y^2+f_N(-x)x^2y+x^6=0
\eqno(1.1)
$$
and symplectic form $\omega=\sum_{i=1}^{2N-2}\delta x(z_i)\wedge
{\delta y\over y}(z_i)$. Here $f_N(x)
=\sum_{i=0}^Nu_ix^i$ is a generic polynomial of degree $N$,
and the parameters $u_i$ can be viewed as the moduli of the
spectral curve. A system with the desired properties can be
obtained as follows. Let $q_n,p_n$ be complex 3-dimensional (column) vectors satisfying 
$q_n^Tp_n=0$ and the reflexivity condition
$p_n=hp_{-n-1}$, $q_n=hq_{-n-1}$, where $h$ is the $3\times 3$
matrix whose only non-zero entries are $h_{31}=h_{22}=h_{13}=1$.
Let $a,b,c$ be $3\times 3$ matrices satisfying
$$
a^2=1,
\quad
ab=ba,
\quad
b^2=ac+ca,
\quad
bc=cb,
\quad
c^2=0.\eqno(1.2)
$$
Consider the dynamical system 
$$
\eqalign{
\dot p_n&=
{p_{n+1}\over p_{n+1}^Tq_n}+{p_{n-1}\over p_{n-1}^Tq_n}+\mu_np_n,
\quad
\dot q_n=
-{q_{n+1}\over p_n^Tq_{n+1}}-{q_{n-1}\over p_n^Tq_{n-1}}-\mu_np_n
\cr
\dot a&=\left\{{q_{m-1}p_m^T
\over p_m^Tq_{m-1}}-{q_mp_{m-1}^T
\over p_{m-1}^Tq_m},b\right\},\quad\quad
\dot b=\left\{{q_{m-1}p_m^T
\over p_m^Tq_{m-1}}-{q_mp_{m-1}^T
\over p_{m-1}^Tq_m},c\right\},\quad\quad
\dot c=0.\cr}
\eqno(1.3)
$$
where $\mu_n(t)$ is an arbitrary scalar function, and we have set
$m=-{N\over 2}+1$ for $N$ even and $m=-{N\over 2}+{1\over 2}$
for $N$ odd. The system (1.3) appears uncoupled, but it will
not be after imposing twisted monodromy conditions. More precisely,
we have

\bigskip
\noindent
{\bf Main Theorem.} {\it Let $x$ be an external parameter, and set $L_n(x)=1+xq_np_n^T$. Then
\hfil\break
{\rm (a)} There are unique $3\times 3$ matrices 
$g_n(x)=a_nx^2+b_nx+c_n$ which satisfy the periodicity condition
$$
g_{n+1}L_{n+N-2}=L_ng_n\eqno(1.4)
$$
for any fixed data $a_r,b_r,c_r,
(p_n,q_n)_{n=r}^{n=r+N-3}$ with the constraint $q_n^Tp_n=0$.
\hfil\break
{\rm (b)} Consider the dynamical system (1.3)
with $a_m=ah, b_m=bh, c_m=ch$. Then the system is 
integrable in the sense that it is equivalent to 
the following Lax equation
$$
\dot L_n=M_{n+1}L_n-L_nM_n\eqno(1.5)
$$
where $M_n(x)$ is the $3\times 3$ matrix defined by
$$
M_n(x)=x\bigg({q_{n-1}p_n^T
\over p_n^Tq_{n-1}}-{q_np_{n-1}^T\over p_{n-1}^Tq_n}\bigg)
\eqno(1.6)
$$
{\rm (c)} The spectral curve 
$\Gamma=\{(x,y);det\big(yI-g_n(x)
L_{n+N-3}(x)\cdots L_n(x)\big)=0\}$ for the 
Lax equation (1.5) is independent of $n$.
It coincides with the 
Landsteiner-Lopez curve (1.1), and the system (1.3) is Hamiltonian with
respect to the symplectic form $\omega$ 
on the reduced phase space $u_N=1$, $u_{N-1}=0$.
The Hamiltonian is $H=u_{N-2}$ .}  

\bigskip

\centerline{\bf II. The LANDSTEINER-LOPEZ CURVE FOR}
\centerline{\bf THE SYMMETRIC
REPRESENTATION}

\bigskip
It is convenient to list here the main geometric properties of the
curve (1.1), which will henceforth be referred to as the LL (Landsteiner-Lopez) curve. 
It admits the following involution
$$
\sigma: (x,y) \rightarrow \left(-x,{x^4\over y}\right).
\eqno(2.1)
$$

\bigskip
\noindent
{\bf Points above $x=\infty$ and $x=0$}

Above $x=\infty$, there are 3 distinct solutions of the LL equation,
given by $y\sim x^N$, $y\sim x^{-N+4}$ and $y\sim x^2$.
The involution $\sigma$ interchanges the first two, leaving the
third one fixed.

Above $x=0$, there are also three solutions, given by
$y\sim x^2$, $y\sim x^4$, and $y\sim 1$. The involution interchanges
the last two points, while the first one is left fixed.
The points $y\sim x^2$ and $y\sim x^4$ cross each other, but they
are not branching points.

\bigskip
\noindent
{\bf Genus of the LL curve}

The Riemann-Hurwitz formula says that the genus $g$
of the LL curve is given by $2g-2=-6+\nu$, where $\nu$
is the number of branching points (for generic moduli,
the branching index is $2$, which we assume).
The branching points correspond to zeroes of $\partial_yR(x,y)$
$$
3y^2+2y f_N(x)+f_N(-x)x^2=0.
$$
To determine their number, we determine the number of poles
of $\partial_yR(x,y)$. These occur at $x=\infty$.
At $x=\infty$, the three solutions $y\sim x^N$,
$y\sim x^2$, and $y\sim x^{-N+4}$ contribute
respectively $2N$, $N+2$, and $N+2$ poles,
for a total of $4N+4$ poles.
Thus there are also $4N+4$ zeroes. At $x=0$,
there are two zeroes $y\sim x^2$ and $y\sim x^4$,
at each of which $\partial_yR(x,y)$ vanishes
of second order. Thus the number $\nu$ of branching points
is given by $\nu=4N+4-2-2=4N$, and consequently
$$
genus(\Gamma)=2N-2.
$$

\bigskip
\noindent
{\bf Genus of the quotient curve}

Let $g_0$ be the genus of $\Gamma_0=\Gamma/\sigma$.
Since $\Gamma$ has two branch points over $\Gamma_0$,
namely $y\sim x^2$ at $x=\infty$ and $y=x^2$ at $x=0$,
the Riemann-Hurwitz formula applies and gives
$2g-2=2(2g_0-2)+2$, from which it follows that
$$
g_0=N-1.\eqno(2.2)
$$

\bigskip
\noindent
{\bf General case}

The LL curve can be seen as a special case of a general family of curves
defined by the equation
$$
R(x,y)\equiv y^3+f_N(x)y^2+g_{N+2}(x)y+r_6(x)=0,
\eqno(2.3)
$$
where $g_{N+2}$ and $r_6(x)$ are polynomials of degree $N+2$ and $6$,
respectively. This family has $2N+11$ moduli. The genus of a curve $\Gamma$
defined by this equation can be found as before. It equals
$$ genus (\Gamma)=2N
\eqno(2.4)
$$

\bigskip

\centerline{\bf III. CONSTRUCTION OF THE SPIN CHAIN}

\bigskip
We give now the proof of the main theorem. Since finding the desirable
integrable model is an essential component of our result, we construct
the model gradually instead of proceeding from its final description.
It is natural to look for a spin chain of $3$-dimensional vectors
with a period of $N+2$ spins, in order to arrive at a spectral curve
of the form (2.3). The difficult steps are to create an involution
of the form (2.1) and to obtain the correct number of degrees of freedom.

\medskip

\noindent
{\bf The spin chain system}

We look for a spin chain system of the form
$$
\psi_{n+1}=L_n(x)\psi_n\eqno(3.1)
$$
with the operators $L_n(x)$ given by
$L_n(x)=1+xq_np_n^T$
where $x$ is an external variable,
and $q_n, p_n$ are 3-dimensional complex vectors satisfying
the condition $q_n^Tp_n=0$. The vectors $q_n$ and $p_n$ should
be viewed as column vectors, so that $q_n^Tp_n$ is a scalar,
while $q_np_n^T$ is a $3\times 3$ matrix.

\bigskip
\noindent
{\bf Twisted monodromy conditions}

The key feature of the construction is the imposition of
suitable twisted boundary conditions. Now the usual periodicity
condition $L_{n+N-2}(x)=L_n(x)$
can be expressed as $TL=LT$, if we define the monodromy
operator to be  
$(T\psi)_n=\psi_{n+N-2}$.
For the Landsteiner-Lopez curve, we require 
a twisted periodicity condition of the form
$$
g_{n+1}(x)L_{n+N-2}(x)=L_n(x)g_n(x)
\eqno(3.2)
$$
with the $g_n(x)$'s suitable $3\times 3$ matrices to be chosen later.
This requires in turn the following more subtle choice of
monodromy operator $T_n(x)$
$$
T_n(x)=g_n(x)\prod_{k=0}^{N-3}L_{k+n}(x)\eqno(3.3)
$$
where, by convention, the indices in the product of the
$L_{k+n}$'s are in decreasing order as we move from left to right.
The twisted periodicity condition (3.2) is then equivalent to 
$$
T_{n+1}L_n=L_nT_n,
\eqno(3.4)
$$ 
which implies that the eigenvalues of $T_n$ are independent of $n$.
We may thus define the spectral curve of the system $L_n$ by
$$
\Gamma=\{(x,y);\ det(yI-T_n(x))=0\}.\eqno(3.5)
$$

\noindent
{\bf Construction of $g_n(x)$}

We look for $g_n(x)$ under the form
$g_n(x)=a_nx^2+b_nx+c_n$,
where $a_n, b_n, c_n$ are $3\times 3$ matrices. The periodicity
condition $g_{n+1}L_{n+N-2}=L_ng_n$ is equivalent to the following
system of equations
$$
\eqalign{c_{n+1}&= c_n\cr
a_{n+1}q_{n+N-2}p_{n+N-2}^T&=q_np_n^Ta_n\cr
a_{n+1}+b_{n+1}q_{n+N-2}p_{n+N-2}^T
&=
a_n+q_np_n^Tb_n\cr
b_{n+1}+c_{n+1}q_{n+N-2}p_{n+N-2}^T
&=b_n+q_np_n^Tc_n\cr}\eqno(3.6)
$$
We claim that this system can be solved completely
in terms of the following
parameters
$$
\eqalign{&a_r,\ b_r,\ c_r\cr
&(p_r,q_r),\cdots,(p_{r+N-3},q_{r+N-3}), \ \ q_n^Tp_n=0,\
r\leq n\leq r+N-3,\cr}\eqno(3.7)
$$
for any choice of initial index $r$. To see this, define $p_{n+N-2}$,
$q_{n+N-2}$ by
$$
p_{n+N-2}^T=p_n^Ta_n,
\ \
q_{n+N-2}=\lambda_n^{-1}a_n^{-1}q_n,\eqno(3.8)
$$
with $\lambda_n$ a scalar yet to be determined.
Then $p_{n+N-2}^Tq_{n+N-2}=\lambda_n^{-1}q_n^Tp_n=0$
and orthogonality is preserved. With $c_n=c_0$
for all $n$ and $p_{N-2+n}, q_{N-2+n}$
defined already as indicated,
the last two equations in (3.6) can be viewed
as recursion relations defining $a_{n+1}$ and $b_{n+1}$.
Our task is to show now that $\lambda_n$ can be chosen so
as to satisfy the second equation in (3.6),
which we rewrite as
$$
\lambda_n^{-1}a_{n+1}a_n^{-1}q_np_n^Ta_n=q_np_n^Ta_n.
$$
Now the recursive equation for $a_{n+1}$ implies that
$$
a_{n+1}a_n^{-1}q_n
+
b_{n+1}\lambda_n^{-1}a_n^{-1}q_np_n^Ta_na_n^{-1}q_n
=
q_n+q_np_n^Tb_na_n^{-1}q_n.
$$
The second term on the left hand side vanishes
since $p_n^Tq_n=0$. Furthermore, the term $p_n^Tb_na_n^{-1}q_n$
on the right hand side is a scalar, so that the preceding
equation implies that $q_n$ is an eigenvector
for the operator $a_{n+1}a_n^{-1}$.
Thus the second equation in (3.6) is satisfied by
choosing $\lambda_n$ to be the corresponding eigenvalue
$$
a_{n+1}a_n^{-1}q_n=\lambda_nq_n,
\ \
\lambda_n=1+p_n^Tb_na_n^{-1}q_n,\eqno(3.9)
$$
completing the recursive construction.

Note that the spectral curve corresponding to generic chain
constructed above has the form (2.3). The dimension of the phase space
equals $D=27+6(N-2)-(N-2)-(N-2)-8=4N+11$,
which is equal to the dimension of the Jacobian bundle over the family
of curves defined by (2.3).
\bigskip
\noindent
{\bf Involution on the spectral curve}

We turn to the task of choosing the twisted monodromy so that the
spectral curve admits the desired involution (2.1).
Recall that the matrix $h$ is given by $h_{ij}=0$, except for
$h_{13}=h_{22}=h_{31}=1$. In particular, $h^2=1$.
Let us impose the following constraints on the spin chain
and the twisted monodromy conditions
$$
p_n=hp_{-n-1},\ \ \ q_n=hq_{-n-1},
\ \ \
g_n(-x)hg_{-n-N+2}(x)h=x^4.\eqno(3.10)
$$
The first two constraints imply
$$
T_n(-x)=g_n(-x)h\left(\prod_{k=0}^{N-3}L^{-1}_{-k-n-1}(x)\right)h=
g_n(-x)hT_{-n-N+2}^{-1}(x)g_{-n-N+2}(x)h
$$
Therefore the last constraint implies that the spectral curve
$\Gamma$ admits the involution $(x,y)\to \left(-x,{x^4\over y}\right)$.
Here we made use of the fact that $L_n(-x)=L_n(x)^{-1}$,
which follows at once from the orthogonality condition
$q_n^Tp_n=0$.
For generic choice of initial index $m$ in (3.10) the second constraint is non-local 
in term of the corresponding parameters (3.7). It becomes
local for a special choice of $m$. Let us assume for simplicity that $N$
is even. Then we choose $m=-N/2+1$. The constraint (3.10) for $n=m$ is
the following equation for $g_m$
$$g_m(-x)hg_m(x)h=x^4$$
Then the last matrix equation in (3.10) is equivalent to the system of equations (1.2) 
for the matrices
$a\equiv a_mh$, $b\equiv b_mh$, and $c\equiv c_mh$.
The last equation in (1.2) implies that $c$ is a
traceless rank one matrix.
Hence, it can be written
in the form
$$
c=\alpha \beta^T,\ \beta^T\alpha=0,
$$
where $\alpha, \beta $ are orthogonal three-dimensional vectors.
The third equation can be solved for $b$ in the form
$$b=\mu\left(a\alpha \beta^T+\alpha \beta^Ta\right), \ \
\mu^2(\beta^Ta\alpha)=1$$
All the equations are satisfied for any $\alpha, \beta$,
and any choice of $a$ such that $a^2=1$. 
We obtain in this way the crucial fact that the dimension 
of the admissible set of initial data $g_m\leftrightarrow
(a_m,b_m,c_m)$ is equal to $8=4+4$. The first term is the 
dimension of matrices $a$ and
the second term is the dimension of orthogonal vectors $\alpha, \beta$
modulo transformation $\alpha\to \kappa \alpha,\ \ \beta\to \kappa^{-1}\beta$.

\bigskip
\noindent
{\bf Degrees of freedom of the system}

The system of vectors $q_n, p_n$ with the orthogonality
constraint has $5(N-2)$
degrees of freedom. The symmetry condition (3.10) reduces it
to ${5\over 2}(N-2)$ (for say, $N$ even).
Now the system has the following gauge invariances
\medskip
\item{$\bullet$} $(q_n, p_n)\to (\mu_nq_n,\mu_n^{-1}p_n)$.
This removes ${1\over 2}(N-2)$ degrees of freedom.
\item{$\bullet$} a global invariance $(q_n,p_n)
\to (W^{-1}q_n,p_n)$ under $3\times 3$ matrices
$W$ satisfying
$Wh=hW$. Such matrices $W$ are of
the form
$$
W=\pmatrix{w_{11} & w_{12} & w_{13}\cr
w_{21} & w_{22} & w_{21}\cr
w_{13} & w_{12} & w_{11} \cr}
$$
Their space is $5$-dimensional. However,
one degree of freedom has already been accounted
for since diagonal matrices are of the form
of the preceding gauge invariance.

\medskip
Thus the total number of gauge invariances is ${1\over 2}(N-2)+4$,
and the number of degrees of freedom for the variables $q_n$, $p_n$
is $2N-8$. A similar counting also produces the same number $2N-8$
of degrees of freedom for the system when $N$ is odd.
Now, as we saw in the previous section,
the system $a_m,b_m,c_m$  has $8$
degrees of freedom.
Altogether, the number of degrees of freedom
of our dynamical system is then
$$
\# degrees\{a_0,b_0,c_0;(q_0,p_0),\cdots (q_{N-3},p_{N-3})\}
=2N,
\eqno(3.11)
$$
which is the same as the dimension of the geometric phase
space constructed out of the curve $\Gamma/\sigma$ and
its Jacobian.

\bigskip
\noindent
{\bf The dynamical equations of motion for $q_n,p_n$}

The equations of motion are determined by the matrix $M_n$
completing $L_n$ into a Lax pair
with equations of motion
$\dot L_n=M_{n+1}L_n-L_nM_n$.
In this case, they are given by
the matrices $M_n(x)$ in (1.6).
We claim that the matrices $M_n$ satisfy the following
periodicity condition
$$
M_{n+N-2}(x)=a_n^{-1}M_n(x)a_n.
\eqno(3.12)
$$
In fact, the periodicity conditions for $q_n$ and $p_n$ imply
that
$$
M_{n+N-2}(x)
=
x\left({a_{n-1}^{-1}q_{n-1}p_n^Ta_n\over p_n^Ta_na_{n-1}^{-1}q_{n-1}}
-
{a_n^{-1}q_np_{n-1}^Ta_{n-1}\over p_{n-1}^Ta_{n-1}a_n^{-1}q_n}\right).
$$
Using the fact that $q_{n-1}$ is an eigenvector of $a_na_{n-1}^{-1}$,
the first term on the right hand side can be easily recognized
as $a_n^{-1}{q_{n-1}p_n^T\over p_n^Tq_{n-1}}a_n$.
Similarly, the second term can also be rewritten as
$a_n^{-1}{q_np_{n-1}^T\over p_{n-1}^Tq_n}a_n$,
using the fact that $p_n^T$ is an eigenvector (on the left)
of the matrix $a_{n}a_{n+1}^{-1}$
$$
p_n^Ta_na_{n+1}^{-1}=\lambda_np_n^T.
\eqno(3.13)
$$
To prove this identity, we use first the recursive relation
defining $a_{n+1}$ and obtain
$$
p_n^Ta_{n+1}
+{1\over\lambda_n}p_n^Tb_{n+1}a_n^{-1}q_np_n^Ta_n
=p_n^Ta_n.
$$
This implies already that $p_n^T$ is an eigenvector
$$
p_n^Ta_na_{n+1}^{-1}
=(1-{p_n^Tb_{n+1}a_n^{-1}q_n\over\lambda_n})^{-1}p_n^T
$$
It remains only to simplify the expression for the eigenvalue.
This is done using the recurrence relation defining $b_{n+1}$
$$
\eqalign{\lambda_n-p_n^Tb_{n+1}a_n^{-1}q_n=&
1+p_n^T(b_n-b_{n+1})a_n^{-1}q_n\cr
=&1+p_n^T({1\over\lambda_n}c_{n+1}a_n^{-1}q_np_n^Ta_n
-q_np_n^Tc_n)a_n^{-1}q_n=1.\cr}
$$
The proof of the relation (3.13) and hence of the periodicity
relations for $M_n$ is complete.

\bigskip
\noindent
{\bf Equations of motion for $a_m,b_m,c_m$}

Let $\Psi_n$ be the solution of the equations
$\Psi_n=L_n\Psi_n,\ \ \partial_t\Psi_n=M_n\Psi_n$,
which is the eigenvector for the monodromy matrix
$$
y\Psi_n=g_n\Psi_{n+N-2}.
$$
Taking the derivative of the last equation
we obtain
$$
\dot g_n=M_ng_n-g_nM_{n+N-2},
\eqno(3.14)
$$
which is equivalent to the equations
$$
\eqalign{\dot a_n&=\left({q_{n-1}p_n^T\over p_n^Tq_{n-1}}
-{q_np_{n-1}^T\over p_{n-1}^Tq_n}\right)b_n-
b_na_n^{-1}\left({q_{n-1}p_n^T\over p_n^Tq_{n-1}}
-{q_np_{n-1}^T\over p_{n-1}^Tq_n}\right)a_n
\cr
\dot b_n&=\left({q_{n-1}p_n^T\over p_n^Tq_{n-1}}
-{q_np_{n-1}^T\over p_{n-1}^Tq_n}\right)c_n-
c_na_n^{-1}\left({q_{n-1}p_n^T\over p_n^Tq_{n-1}}
-{q_np_{n-1}^T\over p_{n-1}^Tq_n}\right)a_n
\cr
\dot c_n&=0.\cr}\eqno(3.15)
$$
The key consistency condition which has to be verified is that
for $m=-{N\over 2}+1$, this dynamical system restrict to the variety
of matrices $a=a_mh,b=b_mh,c=c_mh$ defined by the equation (1.2).
Among these, the difficult equation to check is $b^2=ac+ca$,
and we turn to this next. Here we have assumed to be specific that
$N$ is even. The case of $N$ odd is similar.

\bigskip
\noindent
{\bf Relations at $m=-{N\over 2}+1$}

We claim that the periodicity of the system together with 
their involution relations imply the following relations at
$m=-{N\over 2}+1$
$$
\matrix{
a_m^{-1}q_{m-1}=hq_m\quad\quad
&p_m^Ta_m=p_{m-1}^Th\cr
a_m^{-1}q_m=\lambda_mhq_{m-1}\quad\quad
&\lambda_{m-1}p_{m-1}^Ta_m=p_m^Th\cr}
\eqno(3.16)
$$
To see this, we note that the periodicity conditions with $n=m$
and $n=m-1$ give respectively
$$
\eqalign{
q_{-m}={1\over\lambda_m}a_m^{-1}q_m,\ &
p_{-m}^T=p_m^Ta_m\cr
q_{-m-1}={1\over\lambda_{m-1}}a_{m-1}^{-1}q_{m-1},
\ &
p_{-m-1}^T=p_{m-1}^Ta_{m-1}\cr}
$$
On the other hand, the involution relation with $n=-m$
and $n=-m-1$ give
$$
\eqalign{
&q_{-m}=hq_{m-1},\ \ p_{-m}=hp_{m-1}\cr
&q_{-m-1}=hq_m,\ \ p_{-m-1}=hp_m\cr}
$$
Eliminating $q_{-m}$, $p_{-m}$, $q_{-m-1}$, $p_{-m-1}$
between these relations, and applying the relation
$a_n^{-1}q_n=\lambda_na_{n+1}^{-1}q_n$,
$p_n^Ta_n=\lambda_np_n^Ta_{n+1}$, we obtain the desired
relations.

\medskip
In terms of $a,b,c$, the above relations
imply in particular
$$
p_{m-1}^Ta={1\over\lambda_{m-1}}p_m^T,
\ \
a^{-1}q_{m-1}=q_m.\eqno(3.17)
$$
We now claim that
$$
\lambda_m=\lambda_{m-1}=1.\eqno(3.18)
$$
In fact, the first and third relations in (3.16) imply at once
$q_{m-1}=\lambda_{m-1}(a_mh)^2q_{m-1}$, and hence $\lambda_{m-1}=1$.
Next we show that $\lambda_m=1$.
Recalling the expression
(3.9) for $\lambda_{m-1}$, we may also write
$$
\lambda_{m-1}=1+p_{m-1}^Tb_{m-1}a_{m-1}^{-1}q_{m-1}
=1+p_m^Ta(b_{m-1}h)q_m.
$$
using the facts that $p_{m-1}^T={1\over\lambda_{m-1}}p_m^Ta$ 
and $a_{m-1}^{-1}q_{m-1}=\lambda_{m-1}hq_{m-1}$.
We use now the
inductive relation on the $b_n$'s
$$
b+ca^{-1}q_{m-1}p_{m-1}^Ta=(b_{m-1}h)+q_{m-1}p_{m-1}^Tc.
$$
Substituting in the previous formula for
$\lambda_{m-1}$ gives
$$
\lambda_{m-1}=1+p_m^Ta(b+ca^{-1}q_{m-1}p_{m-1}^Ta-
q_{m-1}p_{m-1}^Tc)q_m
=1+p_m^Tabq_m
$$
since $p_m^Ta$ and $aq_m$ are proportional to $p_{m-1}^T$
and $q_{m-1}$ respectively, and $p_n$ and $q_n$ are
orthogonal. Since we also know that $\lambda_{m-1}=1$,
we deduce that $p_m^Tabq_m=0$.
Now the relation (3.9) applies to
$\lambda_m$ itself, giving
$$
\lambda_m=1+p_m^Tb_ma_m^{-1}q_m=
1+p_m^Tba^{-1}q_m=1+p_m^T
baq_m=1+p_m^Tabq_m
$$
where we have used the equations (1.2) for $a,b,c$.
Since $p_m^Tabq_m$ is known to vanish, 
it follows that $\lambda_m=1$.

\medskip

We can now return to the equations of motion for
$a_m,b_m,c_m$. The equations (3.16) imply
$$
p_{m-1}^Tq_m={1\over\lambda_{m-1}}p_m^Taq_m
={1\over\lambda_{m-1}}p_m^Tq_{m-1}.
\eqno(3.19)
$$
Using (3.16) and (3.19), it is now easy to recast the
equations of motion (3.15) for $a_m,b_m,c_m$ in terms
of the equations of motion (1.2) for $a,b,c$
$$
\dot a=Qb+bQ,\quad\quad
\dot b=Qc+cQ,\quad\quad
Q={q_{n-1}p_n^T\over p_n^Tq_{n-1}}
-{q_np_{n-1}^T\over p_{n-1}^Tq_n}.
$$
Now the compatibility
condition for $b$ is $b^2=ac+ca$, which implies
$\dot bb+b\dot b=\dot a c+c\dot a$.
Substituting in the previous formulas show that this is verified.

\bigskip
\centerline
{\bf IV. THE SYMPLECTIC FORM}

\bigskip
We turn now to the third statement in the main theorem, which concerns
the Hamiltonian structure of our dynamical system. Since the arguments
here are very close to the ones in our earlier work [10],
except for corrections due to the twisted monodromy conditions,
we shall be very brief.

As in [10], our approach is based on the universal symplectic forms obtained in [8,9]
in terms of Lax pairs. Although our main interest is the
symplectic form $\omega$ defined in Section I, there are other symplectic forms
and flows which can be treated at the same stroke. Thus
we define the following symplectic forms $\omega_{(\ell)}$ 
$$
\omega_{(\ell)}={1\over 2}\sum_{\alpha=1}^3{\rm Res}_{P_{\alpha}}
\Omega_{(\ell)},
\eqno(4.1)
$$
where
$$\Omega_{(\ell)}=
\left(<\psi^*_{n+1}(Q) \delta L_n(x)\wedge \delta \psi_n(Q)>_k
+ \psi^*_{k}\left(\delta g_k g_k^{-1}\right)\wedge \delta\psi_{k}\right)
{dx\over x^{\ell}}.
\eqno(4.2)
$$
The various expressions in this equation are defined as follows.
The notation $<f_n>_k$ stands for the sum :
$$
<f_n>_k=\sum_{n=k}^{k+N-3}f_n.
\eqno(4.3)
$$
The expression $\psi_n^*(Q)$ is the dual Baker-Akhiezer function, which is
the row-vector solution of the equation
$$
\psi_{n+1}^*(Q)L_n(z)=\psi_n^*(Q), \ \ \psi_{k+N-2}^*g_k^{-1}(Q)=
y^{-1}\psi_k^*(Q),
\eqno(4.4)
$$
normalized by the condition
$$
\psi_k^*(Q)\psi_k(Q)=1.
\eqno(4.5)
$$
Note that the last term in the definition of the symplectic form
reflects the twisted boundary conditions. As we shall see, that makes the form
independent of the choice of the initial index $n=k$.

We show now that the symplectic form $\omega_{(0)}$ coincides with $\omega$.
In fact, more generally,
$$
\omega_{(\ell)}=-\sum_{i=1}^{2N-2} \delta \ln y(z_i)\wedge
{\delta x\over x^{\ell}}(z_i).
\eqno(4.6)
$$
The expression $\Omega_{(\ell)}$
is a meromorphic differential on the spectral curve $\Gamma$.
Therefore, the sum of its
residues at the punctures $P_{\alpha}$
is equal to the opposite of the sum of the other residues on $\Gamma$.
For $\ell\leq 2$,
the differential $\Omega_{(\ell)}$ is regular at the points situated over
$x=0$, thanks to the normalization (4.5), which insures
that $\delta\psi_n(Q)=O(x)$. Otherwise, it has poles at
the poles $z_i$ of $\psi_n(Q)$ and at the branch points $s_i$,
where we have seen that $\psi_{n+1}^*(Q)$ has poles.
We analyze in turn the residues at each of these two types of poles.

First, we consider the poles $z_i$ of $\psi_n(Q)$. By genericity,
these poles are all distinct and of first order, and we may write
$$
{\rm Res}_{z_i}\Omega_{(\ell)}=\left(<\psi_{n+1}^*\delta L_n\psi_n>_k+
\psi_{k}^*\left(\delta g_k g_{k}^{-1}\right)\psi_{k}\right)
\wedge {\delta x\over x^{\ell}}(z_i).
\eqno(4.7)
$$
The key observation now is that the right hand side can be
rewritten in terms of the monodromy matrix $T_n(x)$.
In fact, the recursive relations $\psi_{n+1}=L_n\psi_n$ and
$\psi_{n+1}^*L_n=\psi_n^*$
imply that
$$
<\psi_{n+1}^*\delta L_n\psi_n>_k+
\psi_{k}\left(\delta g_k g_{k}^{-1}\right)\psi_{k}=
$$
$$=\sum_{n=k}^{k+N-2}\psi_{k+N-2}^*\left(\prod_{p=n+1}^{k+N-2}L_p\right)
\delta L_n\left(\prod_{p=k}^{n-1}L_p\right)\psi_k
+\psi_{k+N-2}\left(g_{k}^{-1}\delta g_k\right)\psi_{k+N-2}=
$$
$$
=\psi_{k+N-2}^*g_{k}^{-1}\delta T_k\psi_k
=\delta \ln y.
$$
In the last equality, we have used the standard formula for the
variation of the eigenvalue of an operator,
$\psi_k^*\delta T_k \psi_k=
\psi_k^*(\delta y)\psi_k$. Altogether, we have found that
$$
{\rm Res}_{z_i}\Omega_{(\ell)}=\delta \ln y(z_i)\wedge {\delta x\over x^{\ell}}(z_i).
$$

The second set of poles of $\Omega_{(\ell)}$ is the set of branching points
$s_i$ of the cover. Arguing as in [10] p. 563, we find
$$
{\rm Res}_{s_i}\Omega_{(\ell)}=
{\rm Res}_{s_i}\left[ <\psi_{n+1}^*\delta L_n d\psi_n>_k+
\psi_{k}^*\left(\delta g_k g_{k}^{-1}\right)d\psi_{k}\right]
\wedge {\delta y \,dx\over x^{\ell}dy}\ .
$$
Due to the identities $dL(s_i)=dg_k(s_i)=0$, this can be rewritten as
$$
{\rm Res}_{s_i}\Omega_{(\ell)}=
{\rm Res}_{s_i}\left[ \left(\psi_{k+N-2}^*g_k^{-1}
\delta T_k d\psi_k\right)
\wedge {\delta y dx\over x^{\ell}dy}\right]\ .
\eqno(4.8)
$$
Next, exploiting the antisymmetry of the wedge product, we may replace $\delta T_k$ in
(4.8) by $(\delta T_k-\delta y)$. Then using the identities
$$
\psi_{k+N-2}^*g_k^{-1}(\delta T_k-\delta y)=
\delta \left(\psi_{k+N-2}^*g_k^{-1}\right) (y-T_k)
$$
$$(y-T_k)d\psi_k =(dT_k-dy)\psi_k,
$$
which result from $\psi_{k+N-2}^*g_k^{-1}(T_k-y)=(T_k-y)\psi_k=0$, we obtain
$$
{\rm Res}_{s_i}\Omega_{(\ell)}={\rm Res}_{s_i}\left(\delta
\left(\psi_{k+N-2}^*g_k^{-1}\right)
(dL-dy)\psi_k\right)\wedge
{\delta y dx\over x^{\ell} dy}
$$
Arguing as before we arrive at
$$
{\rm Res}_{s_i}\Omega_{(\ell)}=
{\rm Res}_{s_i}\left(\psi_{k+N-2}^*g_k^{-1}\delta \psi_k\right)\wedge
 \delta y {dx\over x^{\ell}}.
$$
The differential form
$$
\left(\psi_{k+N-2}^*\delta \psi_k\right)\wedge \delta y {dx\over x^{\ell}}.
$$
is holomorphic at $x=0$ for $0\leq \ell\leq 2$.
Therefore
$$
\sum_{s_i}{\rm Res}_{s_i}\left(\psi_{k+N-2}^*g_k^{-1}\delta \psi_k\right)
\wedge \delta y {dx\over x^{\ell}}=
-\sum_{i=1}^{2N-2}{\rm Res}_{z_i}\left(\psi_{k+N-2}^*g_k^{-1}\delta \psi_k\right)
\wedge \delta y {dx\over x^{\ell}}
$$
Using again the fact that $\psi_{N+k-2}^*g_k^{-1}=y^{-1}\psi_k^*$,
the right hand side of the last equation can be recognized as
$$
\sum_{i=1}^{2N-2}\delta \ln y(z_i)\wedge {\delta x(z_i)\over x^{\ell}(z_i)}.
$$
Finally we obtain
$$
2\omega_{(\ell)}=-\sum_{i=1}^{2N}{\rm Res}_{z_i}\Omega_{(\ell)}-\sum_{s_i}{\rm
Res}_{s_i}\Omega_{(\ell)}=-2
\sum_{i=1}^{2N-2}\delta \ln y(z_i)\wedge {\delta x(z_i)\over x^{\ell}(z_i)}
.
$$
The identity (4.6) is proved.

\bigskip
\noindent
{\bf The Hamiltonian of the Flow}

Let ${\cal M}_{(\ell)}$ be the reduced phase space defined by the following constraints
$$
\eqalign{
{\cal M}_{(0)}&=\{(q_n,p_n;a_n,b_n,c_n);u_{N}=\alpha_0, u_{N-1}=
\alpha_1\}/G\cr
{\cal M}_{(2)}&=\{(q_n,p_n;a_n,b_n,c_n);u_{0}=\alpha_0, u_{1}=\alpha_1\}/G\cr}
$$
where $(q_n,p_n,a_n,b_n,c_n)$ satisfy the conditions of the previous sections,
$G$ is the group of all allowable gauge transformations, and $\alpha_0$, $\alpha_1$
are fixed constants.

\medskip
\noindent
{\bf Lemma.} {\it Let $\ell$ be either $0$ or $2$.
Then the equations (1.3) restricted on ${\cal M}_{(\ell)}$ are
Hamiltonian with respect to the symplectic form $\omega_{(\ell)}$ given
by (4.6). The
Hamiltonians $H_{(\ell)}$ are given by}
$$
H_{(0)}=u_{N-2},\ \ H_{(2)}=\ln u_N
$$

\noindent{\it Proof.}
By definition, a vector field $\partial_t$ on a symplectic
manifold is Hamiltonian, if its contraction $i_{\partial_t}\omega(X)=
\omega(X,\partial_t)$ with the symplectic form
is an exact one-form $\delta H(X)$. The function $H$ is the Hamiltonian
corresponding to the vector field $\partial_t$. Thus
$$
i_{\partial_t}\omega_{(\ell)}={1\over 2}\sum_{\alpha}{\rm Res}_{P_{\alpha}}
\left(<\psi_{n+1}^*\delta L_n\dot\psi_n>_k-
<\psi_{n+1}^*\dot L_n\delta \psi_n>_k+\right.
$$
$$
\left.+\psi_{k}^*\left(\delta g_kg_k^{-1}\right)\dot\psi_k-
\psi_{k}^*\left(\dot g_kg_k^{-1}\right)\delta\psi_k
\right){dx\over x^{\ell}}
$$
Equation of motion for $\dot\psi_n=(M_n+\mu)\psi_n$ implies
$$\sum_{\alpha}{\rm Res}_{P_{\alpha}}
<\psi_{n+1}^*\delta L_n\dot\psi_n>_k{dx\over x^{\ell}}=
\sum_{\alpha}{\rm Res}_{P_{\alpha}}
<\psi_{n+1}^*\delta L_n(M_n+\mu)\psi_n>_k {dx\over x^{\ell}}=
$$
$$
=\sum_{\alpha}{\rm Res}_{P_{\alpha}}
<\psi_{n+1}^*\delta L_n\psi_n>_k {\mu dx\over x^{\ell}}
$$
We used here the equation
$$
\sum_{\alpha}{\rm Res}_{P_{\alpha}}
<\psi_{n+1}^*\delta L_nM_n\psi_n>_k {dx\over x^{\ell}}=0
$$
which is valid because the corresponding differential is holomorphic
everywhere except at the punctures. We will drop similar terms in
all consequent equations.
The equation of motion (1.5) for $L_n$ implies
$$
<\psi^*\dot L \delta \psi_n>_k=
<\psi^*_{n+1}(M_{n+1}L_n-L_nM_n)\delta \psi_n>_k=
$$
$$
=<\psi^*_{n+1}M_{n+1}\delta \psi_{n+1}-\psi_{n}M_n\delta \psi_n>_k-
<\psi_{n+1}^*M_{n+1}\delta L_n\psi_n>_k=
$$
$$
=\psi_{k+N-2}^*M_{k+N-2}\delta \psi_{k+N-2}-\psi_{k}M_k\delta \psi_k
-<\psi_{n+1}^*M_{n+1}\delta L_n\psi_n>_k
$$
Again the last term does not contribute to the sum of residues.

Using the equation of motion for $g_k$ and the equation
$$
y \delta \psi_k=g_k\delta \psi_{k+N-2}+\delta g_k
\psi_{k+N-2}-\delta y \psi_k
$$
we obtain
$$
\psi_{k}^*\left(\dot g_kg_k^{-1}\right)\delta\psi_{k}=
\psi_{k}^*M_k\delta \psi_k-y\psi_{k+N-2}^*M_{k+N-2}g_k^{-1}\delta \psi_{k}=
$$
$$
\psi_{k}^*M_k\delta \psi_k-\psi_{k+N-2}^*M_{k+N-2}\delta \psi_{k+N-2}-
\psi_{k+N-2}^*M_{k+N-2}\left(g_k^{-1}\delta g_k\right)\psi_{k+N-2}
$$
$$
+\psi_{k+N-2}^*M_{k+N-2}\psi_{k+N-2}\delta\ln y.
$$
The last term does not contribute to a sum of the residues due to
the constraints $\delta \ln y =O(x^{-2})$ for $\ell=0$ and
$\delta \ln y =O(x)$ for $\ell=2$.

The expression for $i_{\partial_t}\omega_{(\ell)}$
reduces to
$$
i_{\partial_t}\omega_{(\ell)}={1\over 2}
\sum_{\alpha}{\rm Res}_{P_{\alpha}} \left(<\psi_{n+1}^* \delta L_n\psi_n>_k+
\psi_k\left(\delta g_k g_k^{-1}\right)\psi_k\right)
{\mu(Q,t)dx\over x^{\ell}}=
$$
$$=
{1\over 2}\sum_{\alpha}{\rm Res}_{P_{\alpha}}  \delta(\ln y)
\mu(t,Q){dx\over x^{\ell}}\ .
$$
The proof can now be completed as in [10], p. 567.

\bigskip
\centerline{\bf V. $\theta$-FUNCTION SOLUTIONS}
\bigskip

Since the system (1.3) is completely integrable, we can obtain exact
solutions in terms of $\theta$-functions associated to the spectral curve.
We give these formulas here without details, since their derivation is
entirely similar to the one in [10] pp. 557-560, taking into account 
the twisted monodromy.

Let $\psi_n$ be the Baker-Akhiezer function, which solves the simultaneous
equations
$\psi_{n+1}=L_n\psi_n,\ \partial_t\psi_n=M_n\psi_n$. Its components
$\psi_{n\alpha}$, $1\leq\alpha=3$, are given by
$$
\psi_{n,\alpha}(t,Q)=\phi_{n,\alpha}(t,Q)
\exp{\left(\int_{Q_{\alpha}}^Q nd\Omega_0+td\Omega^+\right)}
$$
$$\phi_{n,\alpha}(t,Q)=
r_{\alpha}(Q){\theta(A(Q)+tU^++nV+Z_{\alpha})\;\theta(Z_0)\over
\theta(A(Q)+Z_{\alpha})\;\theta(tU^++nV+Z_0)}$$
Here $\theta(Z)$ if the Riemann-theta function associated to
the period matrix of the spectral curve; $A(Q)$ is the Abel map;
$V,U^+$ are the vectors of $B$-periods of 
the meromorphic differentials
$d\Omega_0,\ d\Omega^+$ defined by the following requirements.
The differentials $d\Omega_0$ and $d\Omega_+$ have zero $A$-periods,
they are holomorphic outside the two points $P_1,P_3$ above $\infty$ interchanged
by the involution $\sigma$, with $d\Omega_0$ having simple poles and
residues $\pm1$, while $d\Omega^+$ is of the form $d\Omega^+=\pm dx(1+O(x^{-2}))$ at these two points. The $r_{\alpha}(Q_{\beta})$ are 
meromorphic functions satisfying the normalization condition
$r_{\alpha}(Q_{\beta})=\delta_{\alpha\beta}$ and 
the condition that their divisor of poles
$Z_0$ correspond to the inital data $q_n(0),p_n(0)$ of 
the dynamical system.
Let $P_2$ be the point above $\infty$ fixed by the involution,
and let $d\Omega_1$ be the meromorphic form satisfying
$d\Omega_1+d\Omega_1^{\sigma}=d\Omega^+$, where $d\Omega_1^{\sigma}$ is the
image of $d\Omega_1$ under the involution $\sigma$.

The Laurent expansion
of the last factor as $Q\to P_i$ defines constants 
$v_{i\alpha},w_{i\alpha}$,which depend only on the curve
$$
\eqalign{
v_{2\alpha}=\int_{Q_{\alpha}}^{P_2}d\Omega_0,
\quad
v_{i\alpha}={\rm lim}_{x\to P_i}
(\int_{Q_{\alpha}}^xd\Omega_0\mp {\rm ln}\,x)
\quad i=1,3\cr
w_{2\alpha}=\int_{Q_\alpha}^{P_2}d\Omega_1,
\quad
w_{i\alpha}={\rm lim}_{x\to P_i}
(\int_{Q_\alpha}^xd\Omega_1\mp x),
\quad i=1,3\cr}
$$ 
Let $\Phi_n^{(i)}(t)$ be vectors with coordinates
$$\Phi_{n,\alpha}^{(i)}(t)=\phi_{n,\alpha}(t,P_i)
e^{nv_{i\alpha}+tw_{i\alpha}}$$
Then the vector $p_n$ of the spin chain is the unique (up to multiplication; different choices lead to different gauge choices
$\nu_n(t)$ in our dynamical system (1.3))
three-dimensional vector that is orthogonal to $\Phi_n^{(i)}, i=2,3$, i.e.
$$p_n^T\Phi_n^{(2)} =p_n^T\Phi_n^{(3)}=0$$
and the vector $q_n$ is given by the formula
$$q_n={\Phi_n^{(1)}\over p_n^T\Phi_{n-1}^{(1)}}$$

The leading coefficients of the expansion of the Baker-Akhiezer function
provide also the expression for the variables $a_n$. 
In the normalization $c=(c_{ij})$ with $c_{13}=1$
and $c_{ij}=0$ for all other $i,j$,
we find
$$a_n=\widehat \Phi_{N+n-2}\widehat \Phi_n^{-1}
$$
where $\widehat \Phi_n$ is the $(3\times 3)$ matrix with columns $\Phi_n^{(i)}$.
\bigskip

\centerline{\bf ACKNOWLEDGEMENTS}

\bigskip
D.H.P would like to acknowledge the warm hospitality of the National
Center for Theoretical Sciences in Hsin-Chu, Taiwan, and of the
Institut Henri Poincar\'e, Paris, France,
where part of this work was carried out.

\vfill\eject

\centerline{\bf REFERENCES}

\bigskip
\item{[1]} D'Hoker, E. and D.H. Phong,
``Lectures on supersymmetric Yang-Mills theory and integrable
systems", hep-th/9912271

\item{[2]} D'Hoker, E. and D.H. Phong,
``Calogero-Moser Lax pairs with spectral parameter for general Lie algebras",
Nucl. Phys. {\bf B 530} (1998) 537-610, hep-th/9804124

\item{[3]} D'Hoker, E. and D.H. Phong,
``Calogero-Moser and Toda systems for twisted and untwisted affine Lie
algebras", Nucl. Phys. {\bf B 530} (1998) 611-640, hep-th/9804125

\item{[4]} D'Hoker, E. and D.H. Phong,
``Spectral curves for super Yang-Mills with adjoint hypermultiplet
for general Lie algebras", Nucl. Phys. {\bf B 534} (1998) 697-719,
hep-th/9804126

\item{[5]} Donagi, R. and E. Witten,
``Supersymmetric Yang-Mills and integrable systems",
Nucl. Phys. {\bf B 460} (1996) 288-334, hep-th/9510101

\item{[6]} Ennes, I., S. Naculich, H. Rhedin, and H. Schnitzer,
``One-instanton predictions of a Seiberg-Witten curve from M-Theory:
the symmetric case", Int. J. Mod. Phys. {\bf A 14} (1999) 301,
hep-th/9804151

\item{[7]} Gorsky, A., I.M. Krichever, A. Marshakov, A. Mironov, and A. Morozov,
``Integrability and Seiberg-Witten exact solutions", Phys. Lett. {\bf B 355}
(1995) 466, hep-th/9505035

\item{[8]} Krichever, I.M. and D.H. Phong,
``On the integrable geometry of N=2 supersymmetric gauge theories and soliton equations"
J. Differential Geometry {\bf 45} (1997) 445-485, hep-th/9604199

\item{[9]} Krichever, I.M. and D.H. Phong, ``Symplectic forms in the theory of solitons",
{\it Surveys in Differential Geometry IV: Integral Systems}, ed. by C.L. Terng and K. Uhlenbeck,
International Press (1998) 239-313, hep-th/9708170 

\item{[10]} Krichever, I.M. and D.H. Phong,
``Spin chain models with spectral curves from M Theory",
Commun. Math. Phys. {\bf 213} (2000) 539-574,
hep-th/9912180

\item{[11]} Landsteiner, K. and E. Lopez,
``New curves from branes", 
Nucl.Phys. {\bf B 516} (1998) 273-296, hep-th/9708118

\item{[12]} Martinec, E.,
``Integrable structures in supersymmetric gauge and string theory",
Phys.Lett. {\bf B 367} (1996) 91-96, hep-th/9510204

\item{[13]} Martinec, E. and N. Warner,
``Integrable systems and supersymmetric gauge theories",
Nucl. Phys. {\bf B 459} (1996) 97-112, hep-th/9509161

\item{[14]} Seiberg, N. and E. Witten,
``Electro-magnetic duality, monopole condensation,
and confinement in ${\cal N}=2$ supersymmetric Yang-Mills theory",
Nucl. Phys. {\bf B 426} (1994) 19-53, hep-th/9407087

\item{[15]} Seiberg, N. and E. Witten,
``Monopoles, duality, and chiral symmetry breaking
in ${\cal N}=2$ supersymmetric QCD",
Nucl. Phys. {\bf B 431} (1994) 19-53, hep-th/9410167

\end